\documentclass[mathleft
]{an}
\usepackage{amssymb}
\usepackage{graphicx}
\usepackage{times}
\usepackage{color}
\usepackage{natbib}
\bibpunct{(}{)}{;}{a}{}{,}
\begin{document}

\Pagespan{1}{6}
\Yearpublication{0000}%
\Yearsubmission{2014}%
\Month{4}%
\Volume{000}%
\Issue{00}%

\title{Modelling the Magnetic Field Configuration of Neutron Stars}

\author{Riccardo Ciolfi
\thanks{
  \email{riccardo.ciolfi@aei.mpg.de}}
}
\titlerunning{Neutron Star Magnetic Field Configuration}
\authorrunning{R. Ciolfi}
\institute{ Max-Planck-Institut f\"{u}r Gravitationsphysik
(Albert-Einstein-Institut), Am M\"{u}hlenberg 1, 14476 Potsdam,
Germany }

\received{2014 Apr 30} \accepted{2014 May 13} \publonline{2014 Aug 01}

\abstract{The properties of the extremely strong magnetic fields of
  neutron stars affect in a unique way their evolution and the
  associated phenomenology. Due to the lack of constraints from direct
  observations, our understanding of the magnetic field configuration
  in neutron star interiors depends on the progress in theoretical
  modelling. Here we discuss the effort in building models of
  magnetized neutron stars focussing on some of the recent
  results. In particular, we comment on the instability of purely
  poloidal and purely toroidal magnetic field configurations and on
  the evidence in favour of the so-called twisted-torus solutions. We
  conclude with an outlook on the present status of the field and
  future directions.}


\maketitle

\section{Introduction}\label{intro}

One of the most important features characterizing neutron stars (NSs)
is their extremely strong magnetic field,
reaching surface strengths of $\sim10^{13}$~G for ordinary pulsars
and up to $\sim10^{15}$~G in the case of {\it magnetars}
\citep{DuncanThomp92,Mereghetti2008}.
NS interiors harbor even stronger fields, possibly by one order of
magnitude or more.
Such extreme magnetic fields are closely related to the observational
signatures which distinguish NSs from other astrophysical sources,
including, e.g., the continuous pulsar radiation and the spectacular
flaring activity of magnetars.
Moreover, magnetic fields are responsible for structural deformations
associated with gravitational wave (GW) emission
\citep{Bonazzola1996,Cutler2002}
potentially detectable in the near future by ground-based
interferometers such as Advanced Virgo and Advanced LIGO
\citep{Harry2010,Accadia2011}.

The pivotal role played by magnetic fields in the physics and
astrophysics of NSs raises a challenging but also crucial
question for which a satisfactory answer is still missing:
{\it what is the internal magnetic field configuration of a
  neutron star?}
Properties like the amount of magnetic energy stored inside a NS
and its geometrical distribution are well-known to affect
its evolutionary path and consequently its emission properties.
Clear examples come from recent magneto-thermal evolution studies
(\citealt{Vigano2013} and references therein), where it is shown how a
different strength and distribution of magnetic fields (and hence
electric currents) in the stellar interior leads to a very different
phenomenology, potentially explaining most of the differences
between the various classes of NSs (e.g.~it determines the presence or
absence of bursting activity).
To consider a more specific example, knowing the internal magnetic
field energy and geometry is essential in order to understand the
mechanism behind the rare and extremely energetic {\it giant flares}
of magnetars and to predict their occurrence rate.
Additionally, it is necessary to carry out NS
asteroseismology from the quasi-periodic oscillations observed in the
aftermath of these events (e.g.~\citealt{Gabler2012} and references
therein).
As we further discuss in the following (Section~\ref{GW}), the
magnetically-induced GW emission mechanism mentioned above also
depends sensitively on the internal magnetic field geometry.

While the external manifestation of NS magnetic fields
can be constrained by direct observations, the internal
configuration remains essentially unknown and
we can only rely on a collection of indications mostly
based on evolutionary or energetic considerations.
In this context, our understanding can only be improved with the aid
of theoretical modelling, by considering the widest range
of possible ``realistic'' configurations (i.e.,~matching all the basic
expectations) and predicting their imprints on NS observational
properties. This makes it possible to validate or exclude the
different models that have been proposed, guiding us towards a more
and more precise picture of the magnetic fields hidden inside neutron
stars.

At birth, NSs are hot and completely fluid, highly
convective and differentially rotating.
In the very dynamical first stage of their evolution, magnetic fields
can be redistributed and significantly amplified by several mechanisms
including dynamo action, magnetic winding and possibly shear
instabilities like the magnetorotational instability.
In a short timescale, of the order of seconds to minutes, convection
and differential rotation are damped and magnetic fields start
rearranging towards a magnetohydrodynamic (MHD) equilibrium.
During this second stage, the temperature experiences a rapid decrease
and within few hours it drops below $10^9-10^{10}$~K, leading to the
formation of a solid crust and the likely onset of
superfluid/supeconductive phases in the stellar matter. This sets the
beginning of a third stage, in which the magnetized NS evolves through
dissipative processes on much longer timescales ($\tau_{diss}\gtrsim
10^3-10^6$~years, depending on the magnetic field strength).

In the intermediate stage, the star is still fluid and well-described
by ideal MHD (i.e.,~in the limit of infinite electric
conductivity). Magnetic field rearrangement proceeds on
the Alfv\`en timescale, which is inversely proportional to the average
field strength and typically lies in the range $\tau_A \sim
0.01-10$~s.
Since this timescale is much shorter than a few hours, the magnetic
field has enough time to reach
a stable\footnote{Here we do not refer to absolute stability, but only
  to the stability over timescales longer than several Alfv\`en
  timescales.} MHD equilibrium before the
crust forms and superfluid transitions take place.
If this does not happen, the ongoing hydromagnetic readjustment will
efficiently dissipate most of the initial magnetic energy.
The observation of long-lived magnetic fields excludes the
second possibility and is commonly taken as an indication that the
field should have reached already a stable MHD equilibrium at the time
of crust formation.

In this paper we focus our interest on such an equilibrium.
Even if the long-term magneto-thermal
evolution of the NS will slowly alter the magnetic field energy and
geometry, the initial\footnote{In the
  following, ``initial'' will indicate the time of crust formation and
  the beginning of the long-term evolution.} equilibrium configuration
is extremely important
as it dictates the overall energy budget and other fundamental
properties which will remain almost unchanged on timescales $\lesssim
\tau_{diss}$, such as the magnetic helicity and the relative
contribution of the toroidal (azimuthal) and poloidal (meridional)
components of the field.
Therefore, the initial magnetic field represents not only
a key input for long-term evolution studies, but also an important
element for any model describing observed
NS processes in which magnetic fields play an important
role (e.g.,~magnetar flares).
Moreover, there are potentially observable processes associated with
the early evolution of NSs happening on
timescales $\ll \tau_{diss}$ which carry the direct imprint of the
initial field (e.g.~GW emission from newly-born NSs).

The existence of an initial equilibrium allows for a theoretical
description based on stationary configurations and relatively simple
physics (pure fluid matter, uniform rotation, ideal MHD, no
superfluid/superconducting phases, and so on);
furthermore, the requirement of stability over several Alfv\`en
timescales considerably reduces the space of possible
configurations.
The effort in building models of magnetized NSs and the
investigation of their properties (including stability) has
already made important progress in this direction,
although so far no solution has been proven to represent a fully
viable description of the initial NS magnetic field.

In Sections \ref{pol-tor} and \ref{TT}, we give a brief
overview of this work, from the investigation of the simplest field
geometries (already shown to be unstable) to the so-called {\it
  twisted-torus} configuration, which is presently considered a strong
candidate.
As an application, we also discuss the continuous GW emission
of magnetized NSs associated with the structural deformations induced
by magnetic fields, its dependence on the internal field
configuration, and the prospects of detection (Section \ref{GW}).
We conclude with a perspective on open questions and future
directions (Section \ref{outlook}).

\section{Purely poloidal and purely toroidal magnetic field
  configurations}\label{pol-tor}

The growing effort to build equilibrium models of magnetized
NSs started with the early work of \citet{Chandrasekhar1953} and
continued with a large number of analytical and numerical studies
mosty based on the assumpitions of ideal MHD and pure-fluid matter,
well-suited to describe the conditions occurring before the formation
of a solid crust and the onset of superfluidity.

Until the last decade, most attention was devoted to the
simplest magnetic field geometries, consisting of either
a purely poloidal or a purely toroidal field
(e.g. \citealt{Bocquet1995,Frieben2012}).
However, already from the early analytical work on nonrotating
magnetized stars \citep{Markey1973,Tayler1973,Wright1973}, the
stability of these simple geometries was
strongly questioned because of the so-called Tayler (or kink)
instability, which is expected to act on Alfv\`en timescales in both
the purely poloidal and the purely toroidal case.
Recently, the unstable nature of these configurations
was confirmed in Newtonian numerical simulations,
both in the linear regime \citep{Lander:2011b,Lander:2011a} and, for
main-sequence stars, via nonlinear simulations
\citep{Braithwaite2006,Braithwaite2007}.
The analogous system has also been studied by means of nonlinear MHD
simulations in general
relativity (\citealt{Kiuchi2008,Kiuchi2011,Lasky2011,Lasky2012,
Ciolfi2011,Ciolfi2012}; see also \citealt{Ciolfi2014}), further
verifying the presence of a hydrodynamic instability and validating
the basic expectations concerning its onset.

\begin{figure}
\centering
\includegraphics[width=7.cm]{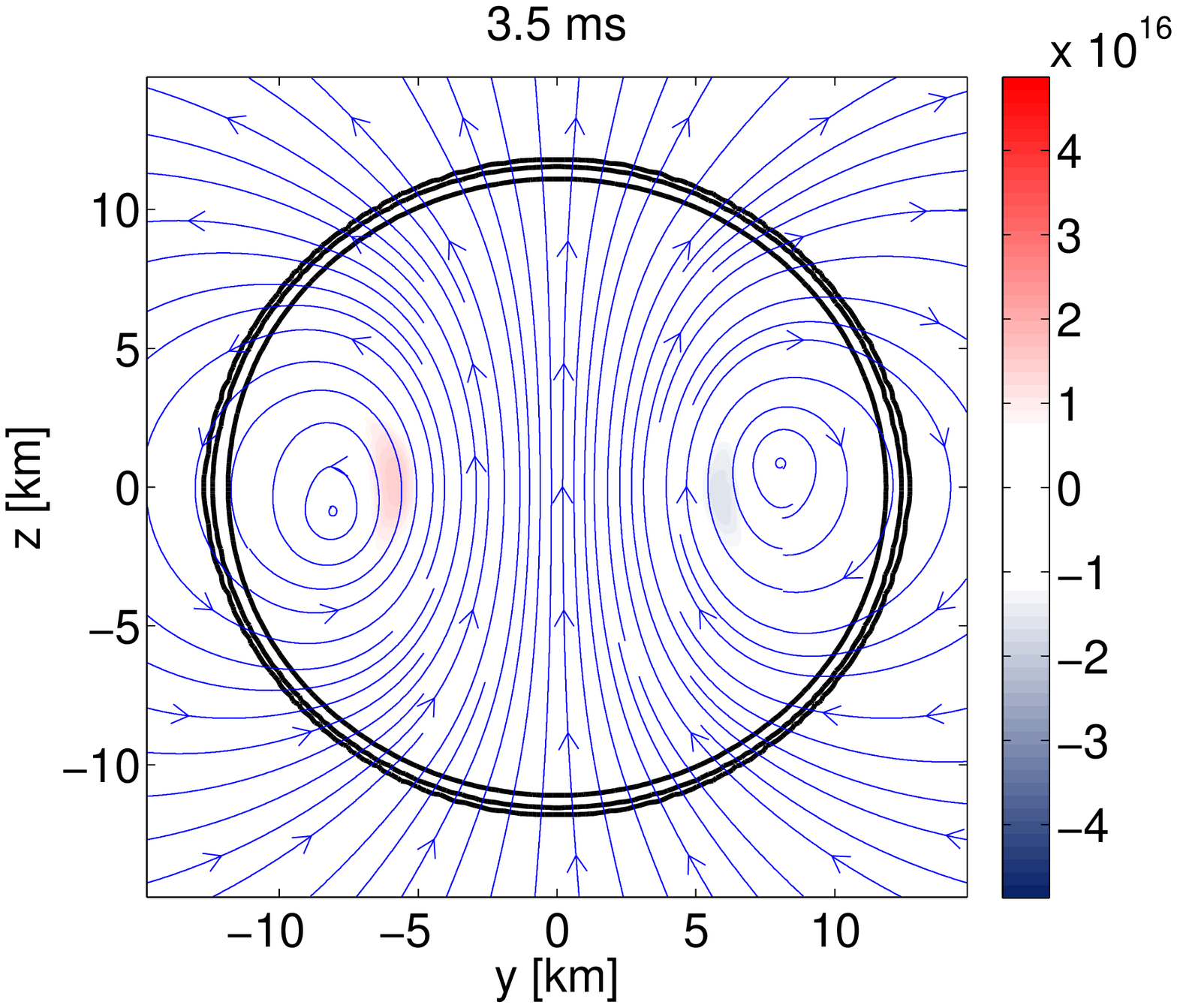}
\\
\vspace{0.3cm}
\includegraphics[width=7.cm]{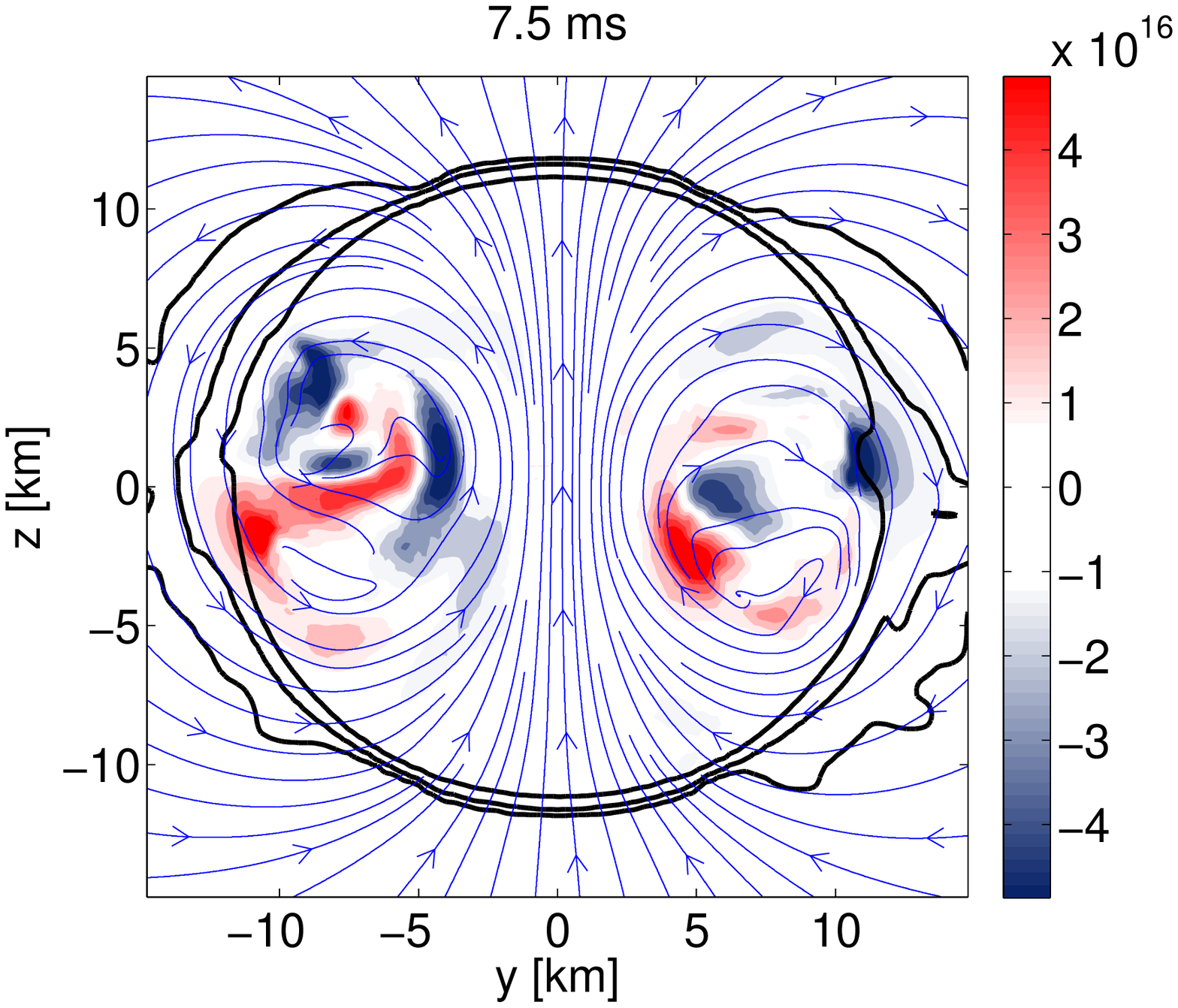}
\caption{Meridional view of the instability of a purely poloidal field
  in a magnetized NS with an initial polar magnetic field strength of
  $6.5\times 10^{16}$~G. Poloidal field lines are shown together with
  the color-coded toroidal magnetic field strength;
  also reported are three rest-mass isodensity contours near the
  stellar surface, corresponding to $(0.02, 0.2, 2) \times
  10^{13}\,{\rm g/cm}^3$. The two frames illustrate respectively the
  saturation of the instability, which sets the beginning af the
  non-linear evolution, and the most violent phase of the field
  rearrangement, when the instability has fully developed. In this
  case the Alfv\`en timescale is $\sim 3$~ms.}
\label{poloidal_inst}
\end{figure}

In Fig.~\ref{poloidal_inst} we show an example of a three-dimensional
general relativistic MHD simulation taken from \citet{Ciolfi2012},
illustrating the instability of a purely poloidal field.
For this geometry, the
Tayler instability is expected to develop in the vicinity of the
neutral line (in the closed field-line region), where the exponential
growth of the initial perturbations is accompanied by the generation
of toroidal fields. Moreover,
the saturation of the instability and the beginning of the nonlinear
rearrangement should take place in about one Alfv\`en timescale, when
the local strength of toroidal fields becomes comparable to the
poloidal one. These predictions are all confirmed in the simulations.
In addition to these confirmations, numerical investigations such as
the one in the example allowed the nonlinear part of the
evolution to be explored, providing hints on the favoured state of the
system.
Although no stable magnetic field configuration was found at the end
of the simulations, the results showed that
a distribution of magnetic energy in poloidal and toroidal fields
close to equipartition and a significant amount of magnetic helicity
emerge naturally in the evolution and thus represent likely features
of a stable geometry \citep{Ciolfi2012}.

General relativistic studies of the poloidal field instability were
also used as a test case to investigate the internal rearrangement
scenario of magnetar giant flares. This scenario suggests that the
violent dynamics associated with a hydromagnetic instability might
represent the mechanism that triggers a giant flare
\citep{ThompsonDuncan1995,ThompsonDuncan2001}.
These studies were able to show that (i) the timescales of the
observed
electromagnetic emission (in particular the intial spike of the flare)
are compatible with an internal rearrangment, which could also
initially trigger a rapid reconfiguration of the external magentic
field in the magnetosphere \citep{Ciolfi2012}, and that (ii) the GW
emission associated with a giant flare event, according to this
scenario, would be hardly detectable in the near future by ground
based interferometers
\citep{Zink2012,Lasky2012,Ciolfi2011,Ciolfi2012}.

For both the purely poloidal and purely toroidal magnetic field
configurations, the presence of rapid (uniform) rotation was long
suggested to have a stabilizing effect, potentially
suppressing the Tayler instability. Nevertheless, the most recent
nonlinear studies found no evidence that rotation can actually stifle
the development of the instability \citep{Lasky2012,Kiuchi2011}.

\section{Mixed-field configurations: the twisted-torus
  geometry}\label{TT}

\begin{figure}
\centering
     \includegraphics[height=4.48cm]{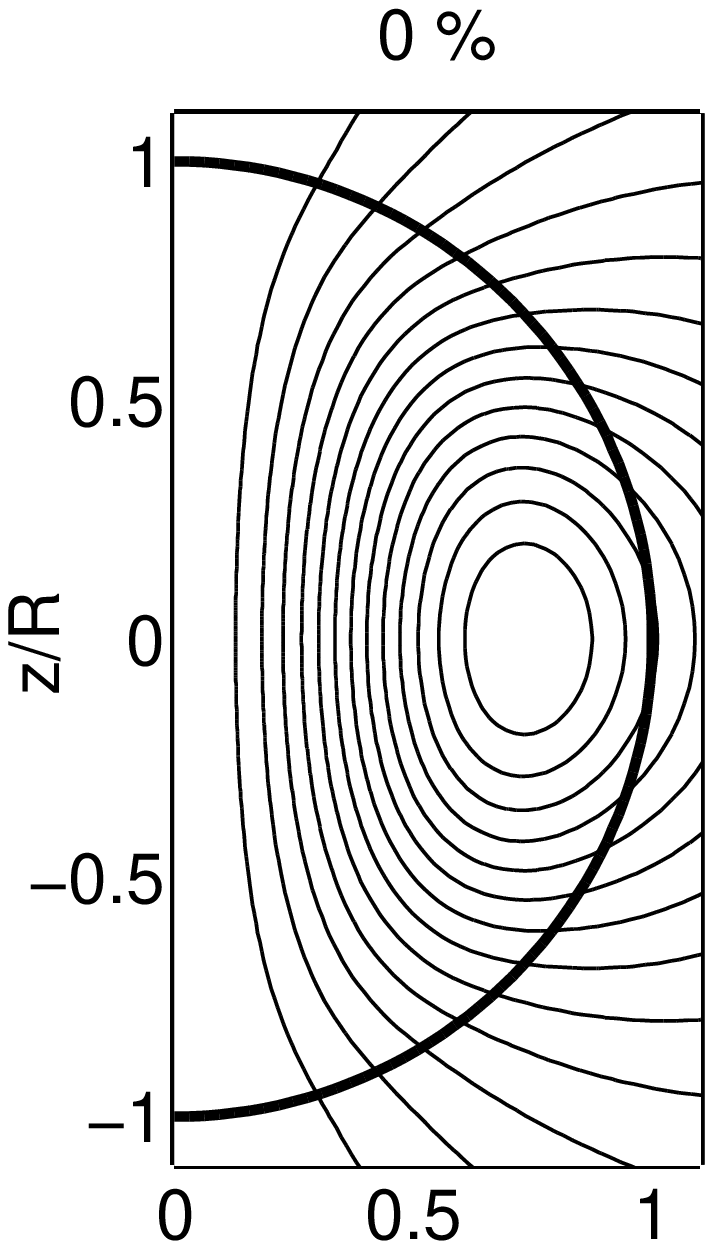}
     \includegraphics[height=4.48cm]{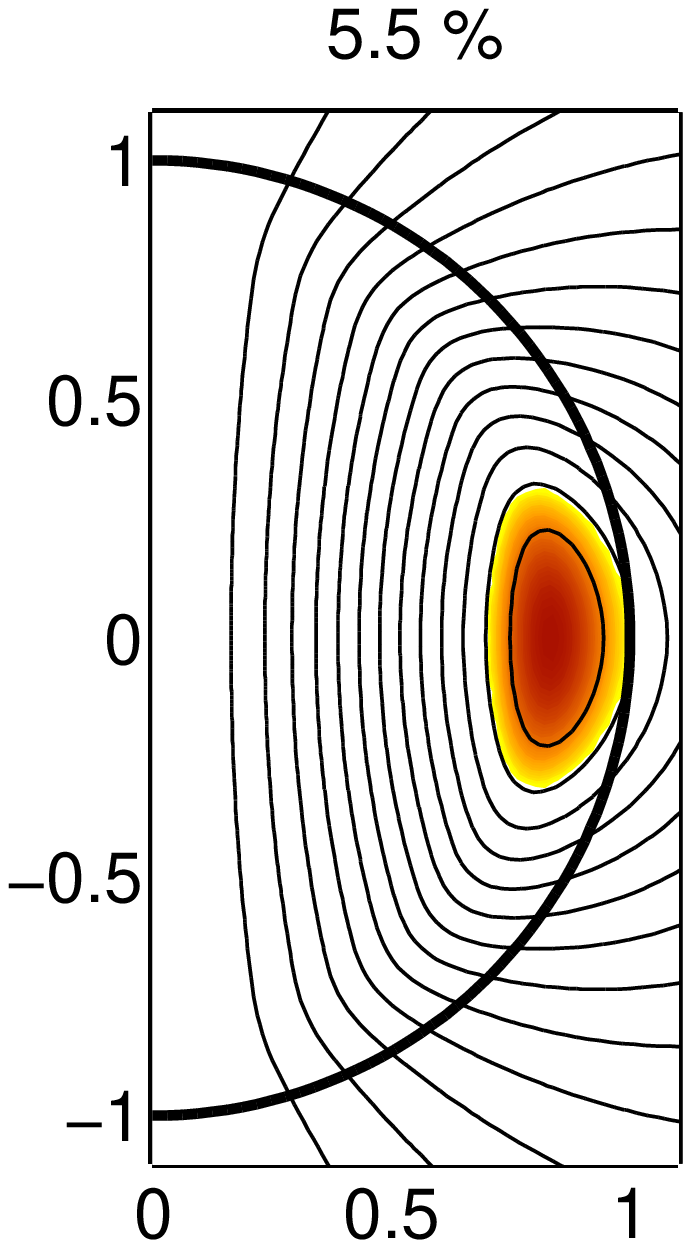}
     \includegraphics[height=4.48cm]{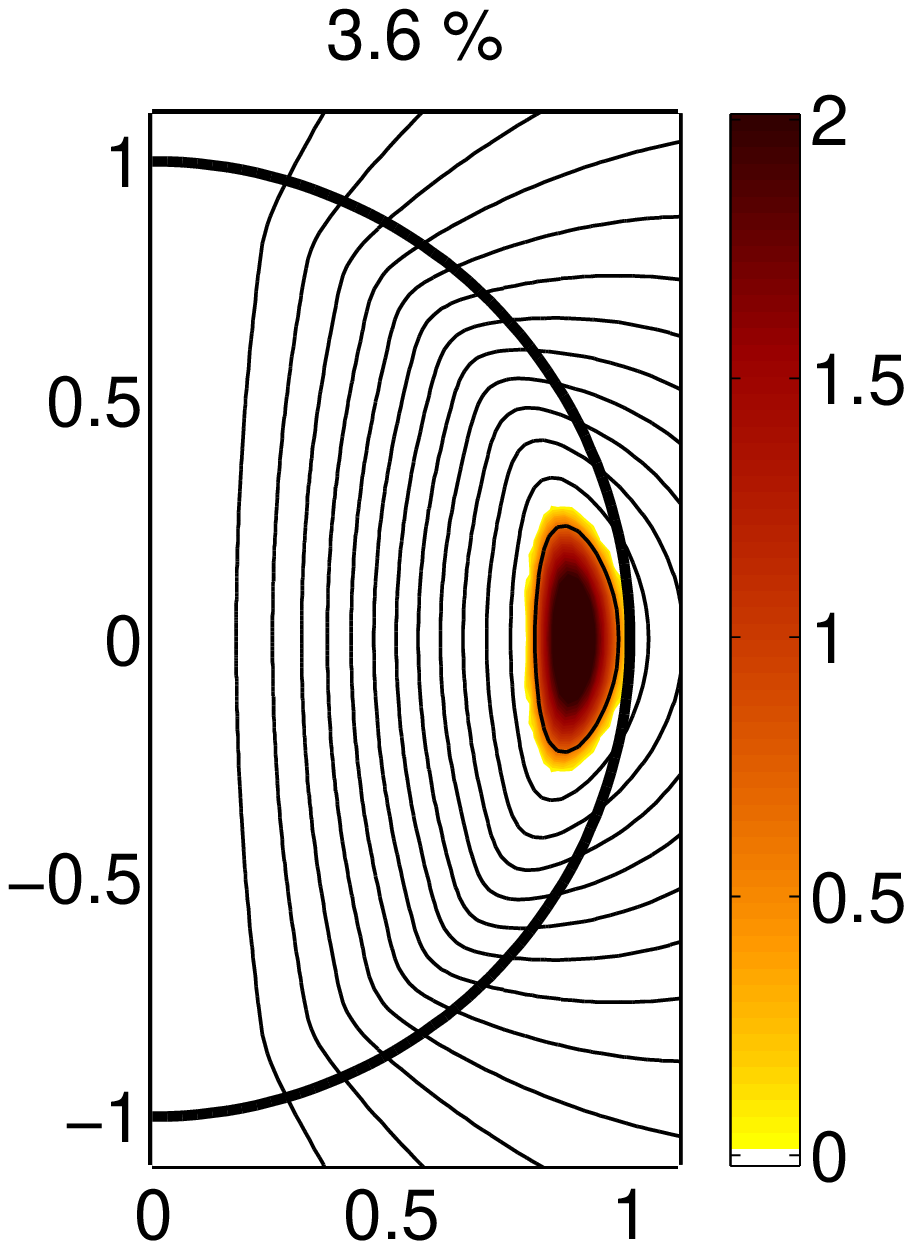}  \\
\vspace{0.2cm}
     \includegraphics[height=4.8cm]{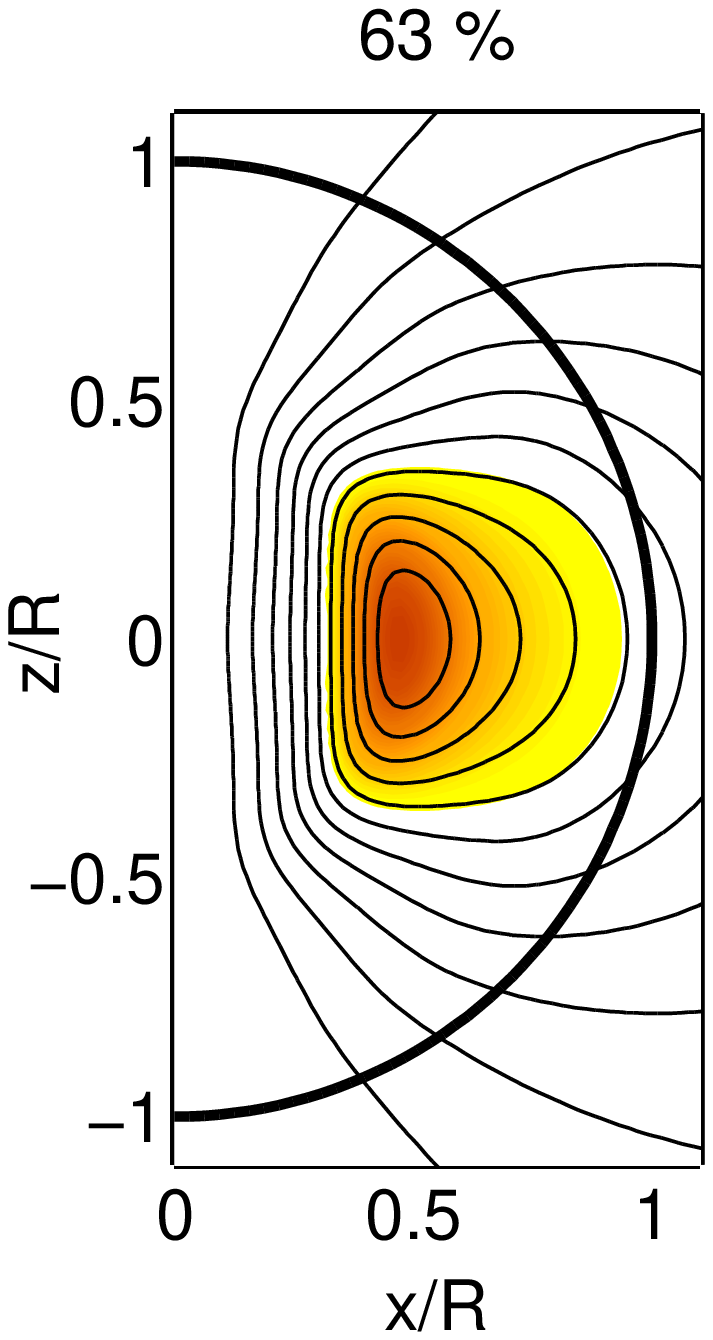}
     \includegraphics[height=4.8cm]{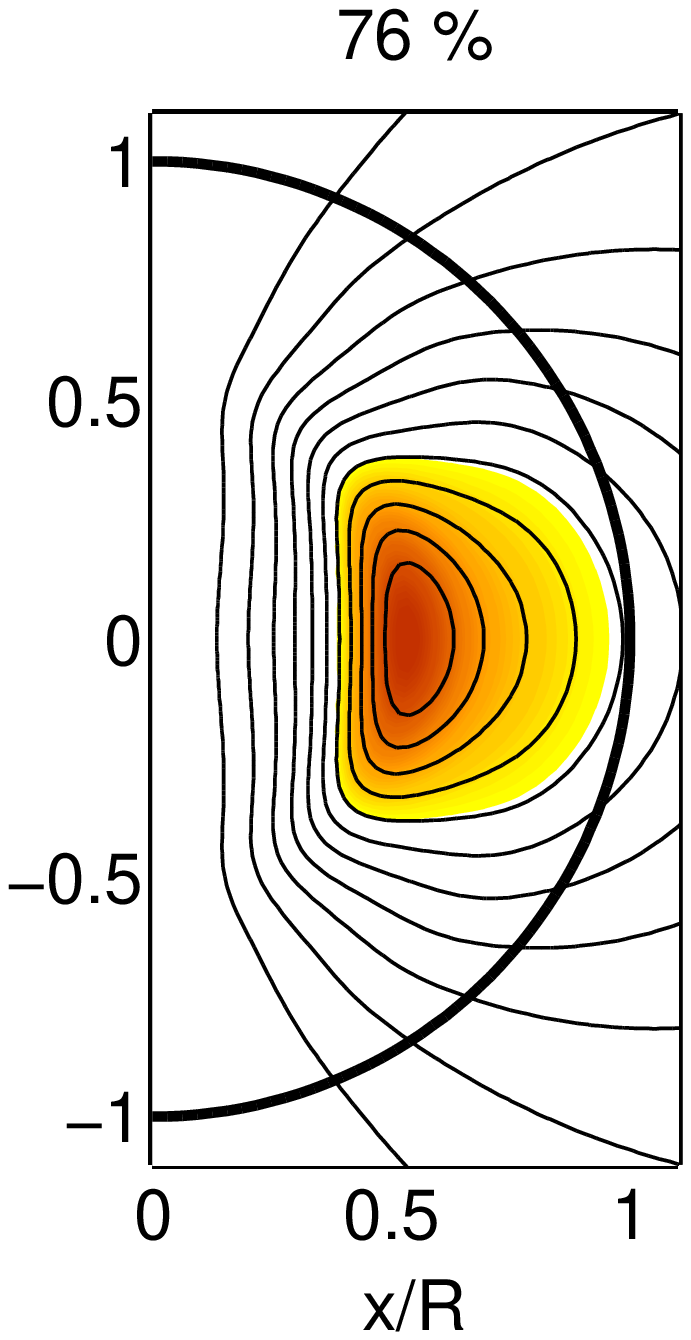}
     \includegraphics[height=4.8cm]{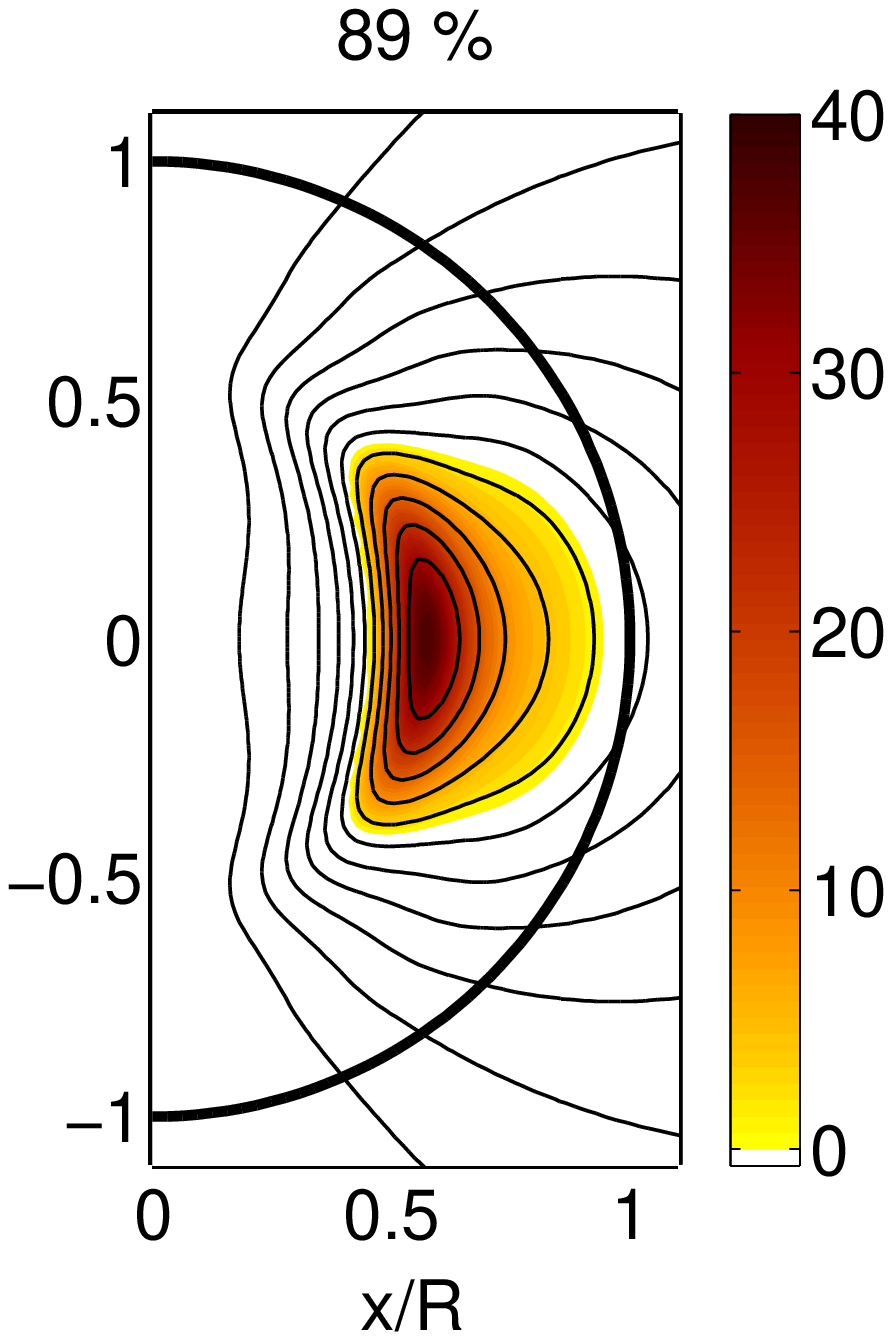}
  \caption{{\it Top panels:} Meridional view of twisted-torus magnetic
    field configurations obtained with the simple commonly-adopted
    current prescription (see \citealt{Ciolfi2013} for
    details). Shown in colour is the toroidal-field strength in units
    of the polar value. $R$ is the stellar radius. The labels on top
    of each panel show the
    toroidal-field energy content relative to the total internal
    magnetic energy ($E_{\mathrm{tor}} /
    E^{\,\mathrm{int}}_{\mathrm{m}}$). {\it
      Bottom panels:} The same as above for example configurations
    obtained with the novel prescription (cf.~Section~\ref{pol-tor} of
    \citealt{Ciolfi2013}).}
\label{fig:TT}
\end{figure}

The accumulated evidence on the instability of purely poloidal and
purely toroidal fields strongly supports the idea that any
long-lived magnetic field configuration in a NS has to consist of
a mixture of poloidal and toroidal field components.
In the present Section, among the possible mixed-field
configurations, we devote our attention to the so-called
\emph{twisted-torus} geometry, which has recently emerged as a
promising candidate for NS interiors.
It consists of an axisymmetric mixed field
where the poloidal component extends throughout the entire star and to
the exterior, while the toroidal one is confined inside the star, in the
torus-shaped region where the poloidal field lines are closed
(cf. Fig.~\ref{fig:TT}).

Compared to other mixed-field configurations, the twisted-torus
has the important feature of having an external poloidal field
(compatible with the observations) while maintaining both the poloidal
and toroidal components continuous at the stellar surface and thus
avoiding the need for surface currents (as, e.g.,~in
\citealt{Colaiuda2008}).
Moreover, a twisted-torus magnetic field appears natural in terms
of the poloidal and toroidal field instabilities: the poloidal
instability takes place in the closed-line region and produces a
stabilizing toroidal component in that region (see
Section~\ref{pol-tor}), while the toroidal field
instability occurs near the symmetry axis and produces a
poloidal field there (see, e.g.,~Fig.~5 in \citealt{Lander2012}).
Apart from these general arguments, important indications in
favour of the twisted-torus configuration came from the Newtonian
simulations performed by \citet{BraithwaiteNordlund}, where this
geometry emerged as the final outcome of the evolution of initial
random fields in a nonrotating fluid star.
Those simulations where adapted to study main-sequence stars, while
the equivalent evidence for NSs and in general relativity is still
missing; nevertheless, this result triggered a growing interest
and in the last decade a number of studies were devoted
to building twisted-torus models of magnetized NSs
both in Newtonian
\citep{Tomimura2005, Yoshida2006, Lander2009, Lander2012,
  Glampedakis2012, Fujisawa2012} and general-relativistic frameworks
\citep{Ciolfi2009,Ciolfi2010,Pili2014}.

All of the twisted-torus models mentioned above gave rather similar
results in terms of the possible configuration of magnetic fields,
despite the different approaches adopted.
Among the common findings, all of these models share the apparently
unavoidable restriction to \emph{poloidal-dominated} geometries,
with an upper limit of $\sim 10\%$ for the toroidal-to-poloidal
magnetic field energy ratio inside the star (unless surface currents
and discontinuous magnetic fields are
included, as in~\citealt{Fujisawa2013}).
For diverse reasons, this limitation is far more serious than it
may appear.
First, a much higher toroidal-field content is expected from the
formation process of magnetized NSs, as a result of strong
differential rotation in the nascent
NS~\citep{ThompsonDuncan1993,Bonanno2003}.
Second, all evidence is that poloidal-dominated geometries are
unstable on Alfv\'en timescales \citep{Braithwaite2009,Lander2012} as
well as purely poloidal geometries, and hence twisted-torus
configurations of this kind may not be realistic.
Moreover, higher toroidal-field energies are needed to
explain the magneto-thermal evolution of highly magnetized NSs, their
bursting activity and their pulse profiles \citep{Pons2011}.

The restriction to poloidal-dominated solutions has recently been
overcome in \citet{Ciolfi2013}, where it has been shown that this
limitation is due to the overly simple
prescription for the azimuthal currents adopted in all the previous
models and that employing a different, more elaborate prescription
allows for any toroidal energy content.
We illustrate this result in Fig.~\ref{fig:TT}, where we
show selected examples among the models presented in
\citet{Ciolfi2013}.
The top panels refer to twisted-torus configurations obtained with the
simple and commonly-adopted prescription. Specifically, we consider
one given model and increase from left to right the local
strength of the toroidal field. In this way, electric
currents get more intense in the outer layers of the NS, the
neutral point is moved outwards and the region of closed-field lines
shrinks.
While the toroidal field strength keeps increasing,
the toroidal magnetic energy content reaches a maximum (in this case
$\sim 5.5$\%) and then starts decreasing again.
Although this shrinking effect cannot be avoided completely, the novel
prescription proposed in \citet{Ciolfi2013} is designed to reduce this
effect and at the same time to allow for larger closed-line regions.
The resulting maximum toroidal energy content can be much higher.
The bottom panels of Fig.~\ref{fig:TT} show configurations with up to
$\sim 90$\% in toroidal energy (i.e.~\emph{toroidal-dominated}),
proving the effectiveness of the suggested recipe.

Being able to build twisted-torus equilibria with any toroidal field
energy content significantly expands the space of known
solutions, including models that are more realistic
compared to the previous poloidal-dominated ones.
In particular, configurations with a relatively high magnetic energy
in the toroidal field are potentially stable over several Alfv\`en
timescales and could thus represent a viable description of the
internal magnetic field of NSs.
Future nonlinear studies, along the lines of what has been done for
the purely poloidal and purely toroidal geometries (see
Section~\ref{pol-tor}), will possibly provide the missing evidence.

An additional finding of \citet{Ciolfi2013} is
that for a fixed polar magnetic field strength, a higher relative
content of toroidal field energy ($> 10\%$) implies in general a
much higher total (poloidal and toroidal) magnetic energy inside the
star.
This means that NSs (and in particular highly magnetized NSs) can
harbor internal magnetic fields that are significantly stronger than
commonly expected\footnote{This can be seen, for instance, in
  Fig.~\ref{fig:TT}, by comparing the color  scale of the top and
  bottom panels, denoting the toroidal field  strength.}, with a
considerable impact on their electromagnetic and GW emission
properties (see the following Section).

\subsection{Continuous GW emission from magnetized NSs}\label{GW}

In the present Section we discuss the GW emission associated
with the magnetically-induced NS deformations, which represents an
ideal example to illustrate the potential impact of the internal
magnetic field configuration on the emission properties of a
magnetized NS.

Magnetic fields alter the NS density and pressure distributions inducing
deformations that can be quantified in terms of the quadrupolar
ellipticity $\epsilon_{_{Q}} \equiv Q/I$, where $Q$ is the
mass-energy quadrupole moment and scales as the magnetic energy
(i.e.~$\propto B^2$), while $I$ is the mean value of the stellar
moment of inertia. Poloidal fields deform
a nonrotating star towards an oblate shape (equatorial radius larger than
polar radius), corresponding to positive $\epsilon_{_{Q}}$, whereas
toroidal fields have the opposite effect; therefore, the sign of the
deformation is reflected in the toroidal-to-poloidal energy ratio. In
general, a magnetized NS that rotates around an axis misaligned with
respect to the magnetic axis and having $\epsilon_{_{Q}} \neq 0$, will
emit a continuous GW signal with amplitude $h \propto |\epsilon_{_{Q}}|
I\Omega^2/d$, where $\Omega$ is the angular velocity and $d$ the
source distance from the observer \citep{Bonazzola1996}.

This emission is most effective for newly-born highly magnetized NSs,
having at the same time the highest spins and the strongest internal
magnetic fields. In this case, the spin down rate is very high and the
relevant timescale for GW emission can be extremely short compared to
the long-term dissipative evolution taking place after the
crust formation. As a consequence, the emitted GWs carry the direct
imprint of the initial magnetic field configuration.
Moreover, since the resulting signal changes significantly over the
time of detection, it has an intermediate nature between a continuous
signal and a long transient.\footnote{The GW signal produced by older
  NSs is genuinely continuous and has the advantage that it does not
  rely on the relatively low rate of birth of highly magnetized NSs
  within reach. On the other hand, it is much weaker and
  only relevant for known pulsars or when considering the GW
  background produced by the entire population of
  magnetized NSs \citep{Marassi 2011,Regimbau2006}.}

Quadrupolar deformations of highly magnetized NSs have been
computed for different magnetic field geometries, including
twisted-torus models
\citep{Lander2009,Ciolfi2010,Ciolfi2013,Haskell2014}.
In particular, for poloidal-dominated twisted-torus configurations and
magnetar-like polar field strengths ($\sim 10^{15}$~G), the
ellipticities are always positive (oblate shape) and lie in the range
$10^{-6}-10^{-5}$ \citep{Lander2009,Ciolfi2010}.
Simple estimates \citep{Gualtieri2011} suggest that newly-born NSs
with such deformations and initial spin periods of the order of
$\sim$ms should be marginally detectable up to distances of the order 
of $\sim$10~kpc (i.e., within our Galaxy) by the advanced ground-based 
interferometers Virgo and LIGO, and up to $\sim 0.1-1$~Mpc by future 
generation detectors such as the Einstein Telescope \citep{Punturo2010}. 
The main limitation for the actual detection of these GW signals comes 
from the relatively low birth rate of highly magnetized NSs, which is 
limited to only a few per century in our Galaxy.

The prospects of detection may increase significantly
for twisted-torus models with a higher toroidal field energy content,
which can harbor much higher internal magnetic
energies. For instance, models with 50\% of the
internal magnetic energy in toroidal fields are found to have
up to a factor of $\sim 5-10$ larger ellipticities
\citep{Ciolfi2013,Haskell2014}.
Since the GW amplitude is proportional to $\epsilon_{_{Q}}$ and
inversely proportional to the source distance, an ellipticity larger
by a given factor $\chi$ will enlarge by the same factor the maximum
distance at which a highly magnetized newly-born NS can be detected.
If we take, for instance, $\chi\sim 10$, the typical distance of
detectability for the Einstein Telescope can expand up to the Virgo 
Cluster scale ($\sim 20$~Mpc), with a corresponding event rate of 
more than one per year.

In addition, magentized NSs with a toroidal energy content $\gtrsim
50$\% are deformed into a prolate shape (negative ellipticity). In
this case, a `spin-flip' mechanism driven by internal viscosity may
occur, leading to an increase in the angle between the spin and
magnetic axes, towards a nearly orthogonal configuration
\citep{Jones1975,Cutler2002}. If this mechanism is effective within
the spin-down timescale, the GW emission is further enhanced, leading
to very optimistic prospects of detection \citep{Stella2005}.

\section{Outlook}\label{outlook}

The effort devoted to build models of magnetized NSs and to study how
magnetic fields affect their observational properties is essential in
order to improve our understanding of the physics and astrophysics of
these objects.
As discussed in the previous Sections, important constraints on the
unknown internal magnetic field configuration of NSs can be obtained
by looking for equilibrium solutions that represent a viable
description of the NS magnetic field at the time of crust
formation, shortly after NS birth and before the long-term dissipative
evolution takes place.
Viable solutions have to satisfy stringent requirements, including
stability over timescales much longer than the Alfv\`en timescale,
which already allowed the exclusion of purely poloidal and
purely toroidal geometries.

Presently, twisted-torus configurations with a substantial amount of
magnetic energy in the toroidal component appear as promising
candidates for stability, although a proper confirmation is still
missing. Future studies will provide a final answer on the viability
of twisted-torus magnetic fields and possibly suggest
alternative solutions to this long-standing open issue.
State-of-the-art nonlinear simulations of magnetized NSs in
general relativity will play an important role in this investigation.

In order to improve the realism of the simulations, various aspects
need to be refined, including a proper treatment of magnetic
fields in the magnetosphere surrounding the star and the use of
realistic equations of state.
Moreover, stable stratification due to composition gradients
is an additional element to be taken into account that might
represent a necessary ingredient for stability
\citep{Reisenegger2009,Mitchell2013}.

As a final note, the onset of superconductivity in newly-born NSs with
internal magnetic fields below $\sim 10^{16}$~G (i.e., ordinary NSs
and possibly some magnetars) can significantly alter the equilibrium
of the system \citep{Lander2014}.
This might happen on a timescale comparable to the formation of a solid
crust -- causing a rapid readjustment of the initial magnetic field
configuration -- or on much longer timescales, producing notable effects
during the long-term evolution of the NS. This is an important aspect
and deserves futher investigation.


\bigskip
\acknowledgements

We thank A. Harte for useful discussions. RC is supported by the
Alexander von Humboldt Foundation.



\end{document}